\documentclass[superscriptaddress,showpacs,preprintnumbers,amsmath,amssymb,floatfix,prd]{revtex4}

\usepackage{graphicx}
\usepackage{dcolumn}
\usepackage{bm}
\usepackage{hyperref}
\hypersetup{colorlinks,citecolor=blue}

\begin{document}

\title{Shadows of Lorentzian traversable wormholes}

\author{Farook Rahaman}
\email{rahaman@associates.iucaa.in}
\affiliation{Department of Mathematics, Jadavpur University, Kolkata 700032, West Bengal,
India}

\author{Ksh. Newton Singh}%
\email[Email:]{ntnphy@gmail.com}
\affiliation{Department of Mathematics, Jadavpur University, Kolkata 700032, West Bengal, India}
\affiliation{Department of Physics, National Defence Academy, Khadakwasla, Pune-411023, India}

\author{Rajibul Shaikh}
\email{rshaikh@iitk.ac.in}
\affiliation{Department of Physics, Indian Institute of Technology, Kanpur, Uttar Pradesh}

\author{Tuhina Manna}
\email{tuhina.manna@sxccal.edu}
\affiliation{Department of Mathematics, St.Xavier's College(Autonomous), 30 Mother Teresa Sarani, Kolkata-700016,
India}

\author{Somi Aktar}
\email{somiaktar9@gmail.com}
\affiliation{Department of Mathematics, Jadavpur University, Kolkata 700032, West Bengal, India}

\date{\today}

\begin{abstract}

The prospect of identifying wormholes by investigating  the shadows of wormholes constitute a foremost source of insight into the evolution of compact objects and it is one of the essential problems in contemporary astrophysics. The nature of the compact objects (wormholes) plays a crucial role on shadow effect, which actually arises during  the strong gravitational lensing. Current Event Horizon Telescope observations have  inspired scientists to study and to construct the shadow images of the wormholes. In this work, we explore the shadow cast by a certain class of rotating  wormhole. To search this, we first compose the null geodesics and study the effects of the parameters on the photon orbit. We have exposed the form and size of the wormhole shadow and have  found that it is slanted as well as can be altered depending on  the  different parameters present in the wormhole spacetime. We also constrain the size and the spin of the wormhole using the results from M87* observation, by investigating the average diameter of the wormhole as well  as  deviation from circularity with respect to the wormhole throat size. In a future  observation, this type of  study may help to indicate the presence of a wormhole in a galactic region.

\end{abstract}

\maketitle

\section{Introduction:}
Invisible astrophysical objects, like wormholes  are still a matter of disquiet  since the discovery of geometrical theory of gravity.  Wormholes are  topologically non-trivial structures of the spacetime connecting our Universe with other universes. The original idea was known as the Einstein-Rosen bridge \cite{Einstein}, and  was more of a mathematical concept, since it did not deal with traversable wormholes. Later Moris and Thorne \cite{kg14} developed the idea of traversable wormholes containing exotic matter in their throat. More developments were carried out by various other researchers \cite{Gon,Ellis,kg1}. In observational astrophysics, wormholes are still a viable subject.  Observational detection means one has to interact with the geometrical structure of the wormhole by which data is acquired. Therefore, scientists  are trying to envisage wormholes in order to acquire perception about its geometrical features. Apart from gravitational lensing  effects \cite{GL1},  another  possible method for probing wormholes is the usage of shadows (a dark region over a luminous contextual). One can recognize the shadow of a wormhole as the result of strong gravitational lensing generating a dark spot in the sky of a remote  viewer. It is actually an optical look exhibited by the wormhole when there is a source of light from stars, accretion flows etc. around it. This manifestation has been investigated  theoretically as well as observationally, by the application of very large baseline interferometry (VLBI) \cite{Doe} techniques at sub-mm wavelengths, for studying the shadow cast by supermassive objects, including the ongoing ambitious project involving Sagittarius A* in our galaxy center \cite{Brod2}. Being at sub mm wavelength this technique significantly reduces the interstellar scattering, and  may allow us to see beyond the compact synchrotron emitting region around SgrA*, since this region is expected to become optically thin at sub-mm wavelengths. Also, VLBI experiments can reach a resolution of the order of the angular size of the gravitational radius of SgrA*. Thus the spacetime geometry could be determined through the analysis of shadow in the vicinity of photon sphere.  Similar ambitious projects include the Japanese VLBI Network (JVN), the Chinese VLBI Network (CVN), the Korean VLBI Network (KVN), the Chinese Space VLBI project \cite{VLBI}.   \par

Compact Objects can cause extreme local deflections of light and can bend light by an  arbitrary large angle, causing the formation of light rings which are circular photon spheres, possessing distinct phenomenological signatures in terms of both electromagnetic and gravitational waves. And since this shadow outline depends on the gravitational lensing of nearby radiation,  its proper surveillance can provide  evidence about the parameters as well as spacetime geometry around the wormhole. Light propagation  i.e.  null geodesics in  compact object (wormhole) spacetime can be perceived  in two ways: trajectories which escape to distant observer  and such that are apprehended by the compact object.  Therefore, a black region appears in the viewer's sky, creating the shadow. The boundary of the shadow can be resolved by the wormhole spacetime metric itself, as it relates to the appearing shape of the photon detained  orbits as observed  by a remote viewer.  Theoretically, although we seek to find an analytical closed form of the shadow edge, it is rarely possible to integrate the geodesic motion. In such cases the null geodesic equations have to be solved numerically. An efficient procedure involves a method called backwards ray-tracing, which requires the propagation of light rays from the observer backward in time and identifying their origin \cite{Backtracing}, instead of evolving the light rays directly from a light source and detect the ones that reach the observer.

Ever since the remarkable image of the shadow of black holes, M87* at the center of the Virgo A galaxy was taken by Event Horizon Telescope in April 2019 \cite{telescope,telescope2}, a new era of research on shadows of dark compact objects like black holes and wormholes have emerged. Following this, in 2020, Kruglov \cite{BS6} considered rational nonlinear electrodynamics and interpreted the super-massive M87* black hole as a regular (without singularities) magnetized black hole.   The shadow observed from the center of M87* was consistent with the shadow  generated through numerical simulation of a Kerr black hole of mass $(6.5\pm 0.7)\time 10^{9}M_\odot$ \cite{telescope}. This shadow can provide us  with the properties of the black hole such as mass, spin, accretion disk and to verify the constraints on fundamental physics. The EHT data strongly suggest that the dimensionless spin parameter $a*$ is $\gtrsim 0.5$. This results in the non-existence of ultra-light boson of mass $\mu_b\sim 10^{-21}$eV, which was considered as a dark matter candidate in the distribution scales of $\sim$ kpc \cite{dav}. Vagnozzi and Visinelli \cite{vagn} used the M87* data to constraint the AdS$_5$ curvature radius in Randall-Sundrum AdS${_5}$ brane-world.

Since in classical theory these black holes are not a source of radiation, the shadow actually targets exactly this feature, for lack of radiation. The shadow is an important observational evidence of the most standard problem in general relativity: the motion of light around compact objects. Earliest works on static Schwarzschild black hole shadows are done by \cite{Syn} and \cite{Luminet}. The influence of charge on the measurements of shadow in case of Riessner-Nordstr\"{o}m black hole were investigated by Zakharov et.al. \cite{Zakharov1,Zakharov2}. Detailed study on rotating Kerr black hole were conducted by \cite{Bardeen}-\cite{Brod} to mention a few. Other pioneering works on black hole shadows include \cite{Yumoto}-\cite{sha12}. The challenge of investigating the spacetime in the vicinity of the event horizon of a black hole, with angular resolution comparable to the event horizon, is almost a thing of the past. In fact direct detection of dynamics of matter at light speed and at strong gravity near a compact object is now almost possible. Very recently in 2020, shadow, quasinormal modes, and quasiperiodic oscillations of rotating Kaluza-Klein black holes were investigated by Ghasemi-Nodehi et.al. \cite{BS1}, while Kala et al. \cite{BS2} studied the deflection of light and shadow cast by a dual-charged stringy black hole. Long et al. \cite{BS3} shown that the shadow size of disformal Kerr black hole in non-stealth rotating solutions in quadratic degenerate higher-order-scalar-tensor (DHOST) increases with a deformation parameter and the shape on both deformation and spin parameter. Quasi-normal modes and shadows of Schwarzschild and high-dimensional Einstein-Yang-Mills spacetimes were also discussed in \cite{BS5}-\cite{BS11}.

While black holes continues to attract much attention, horizonless compact objects are fast gaining popularity recently in the literature for a number of reasons (for a review of the topic, see the recent work of Cardoso and Pani \cite{cardoso}). One important class of horizonless objects are wormholes. Imaging the shadow of a wormhole may help  to progress observational investigation of differentiating the wormhole amongst other compact objects  e.g. black holes, neutron stars etc. In an interesting paper in 2013, Bambi \cite{Bambi_2013} pointed out that there is insufficient evidence to confirm that the supermassive objects {at the center} of many galaxies, including our own, are in fact Kerr black holes. Instead exotic compact objects like wormholes might have been formed in the early universe. Though an exhaustive list of papers on wormhole shadows is not possible, yet \cite{Sarbach}-\cite{Vincent} are many notable works. More recently, Amir et. al. \cite{WS1} worked on the shadows of Kerr like wormholes and charged wormholes \cite{WS2} in Einstein-Maxwell-dilaton theory. Shadows of rotating wormholes were also investigated by Gyulchev et.al. \cite{WS3} , Shaikh \cite{WS4} and Abdujabbarov \cite{WS5}. Wielgus et al. \cite{BS4} constructed Reissner-Nordstr$\ddot{o}$m (RN) wormhole by joining two different RN spacetime i.e. two different mass and charge. This kind of wormholes are not mirror-symmetric which affects the photon sphere and may have two photon rings to a distant observer. This doubling of photon ring was explained by doubling of effective potential maximum. However, mirror-symmetric wormholes can also have multiple photon rings and can exhibit interesting strong lensing effect (\cite{RS2019,RS2019b,NT2021}). These recent ventures \cite{WS6} on the shadow of compact objects have  inspired us to construct the shadow of  wormholes,  as well as,  analyse the shape of the shadows.

The paper is organized as follows: in section II we have described the   motion of photons around   wormholes. The geometric equations defining its shadow are obtained in section III.
Section VI is dedicated to final observations.

\begin{figure*}
\includegraphics[scale=0.43]{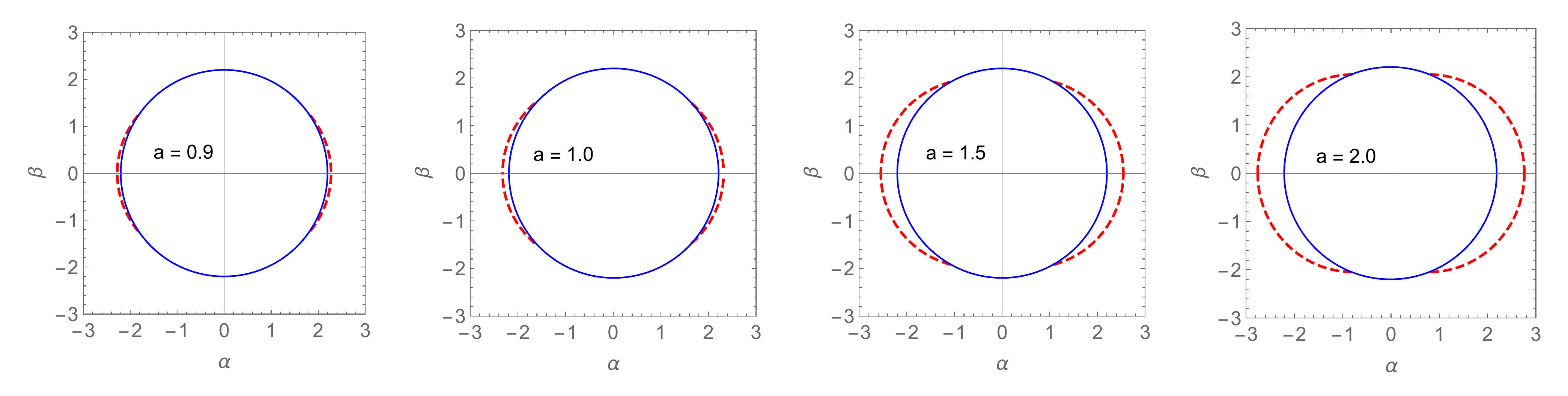}
\caption{Schematic diagram of ergosphere. The region between the ergosurface (red-dashed curve) and the wormhole throat (blue circle) is the ergoregion.}
\label{f1}
\end{figure*}

\section{ The general rotating wormhole spacetime}

Let us consider a general stationary, axisymmetric spacetime  metric, describing a Teo class rotating traversable wormhole \cite{Teo} in Boyer-Linquist coordinates $ (t, r, \theta, \phi) $   as

\begin{eqnarray}
ds^2 &=& -N^2dt^2+\left(1-\frac{b}{r}\right)^{-1}dr^2+r^2K^2
  \left[d\theta^2+\sin^2\theta(d\phi-\omega~ dt)^2\right] , \label{e1}
\end{eqnarray}
where the  spherical polar coordinates   $t, r, \theta$ and $\phi$    are defined by $ -\infty  < t < \infty$, and  $r_0 \leq r < \infty,~ 0 \leq \theta \leq \pi$  and $0 \leq \phi \leq 2\pi $  and in  general, the metric functions  $ N, b, K $  and  $\omega$ depend only on $r$ and  $\theta$. The orbits of the timelike and the spacelike Killing fields ($t^\mu \partial_\mu = \partial_t$ and $\phi^\mu \partial_\mu = \partial_\phi$ respectively ) represent in terms of the parameters $t$ and  $ \phi$.  This metric  is actually  a generalized rotating  form of the static Morris-Thorne wormhole metric \cite{kg14}. Therefore, the metric functions must have some restrictions, which lead to the specific geometry and significant properties of the wormhole. The function $N$ which  regulates the gravitational redshift is often called a redshift function. In order for the wormhole to be traversable,  one has to  {ensure} that there are no curvature singularities or event horizons. Therefore,  the redshift function $N$ should be nonzero and finite all over the region, i.e. from the throat to the spatial infinity. The function $b$ is known as the  shape function  which  regulates the shape of the wormhole.  It is always non-negative and possesses an apparent singularity at $r =  r_{0}= b \geq 0 $, which is  associated with  the throat of the wormhole. The function $b$ should be independent of $\theta$  to avoid any curvature singularity at the throat i.e. it obeys the restriction $\frac{\partial b(r, \theta) }{ \partial \theta} =0$. An important constraint is the flare-out condition at the throat: $ \partial_r b|_{r=r_{0}} < 1$, while $ b(r) < r$  near the throat. Function $K (r, \theta)$  is a regular, positive and non-decreasing function,  defining the proper radial distance $R = rK$, measured at $ (r, \theta$)  from the origin. Actually it   regulates the area radius given by $R$. The function $\omega$ is associated with the angular velocity of the wormhole. Also to confirm that metric \eqref{e1} is regular on the rotation  axis  $ \theta  = 0,  ~\pi $,  the derivatives of $N$ and  $K$   with respect to $\theta$  should vanish on it, i.e. $\frac{\partial N(r, \theta) }{ \partial \theta} =0$, $\frac{\partial K(r, \theta) }{ \partial \theta} =0$. Although, a wide variety of choices for the metric functions  $N, K, b$  and $ \omega$    {can be freely adopted, we} emphasize that all the  solutions do not correspond to wormhole physics.  One  {must} choose such  metric functions   which satisfy the above-mentioned regularity conditions. That class of solutions are the certain cases  of the rotating Teo wormhole. In this study we will consider the specific type  of solutions, where entire  metric coefficients  are functions  of   radial coordinate r only. These solutions reduce to the  static  wormhole   for  zero rotation, i.e. when $\omega = 0$.

Recently, Konoplya and Zhidenko \citep{RK2016} studied quasinormal rings of a wormhole, which  rings like the Schwarzschild or the Kerr black holes and its metric is similar to \eqref{e1}. The specific form of $g_{tt}$ and the shape function $b(r)$ are given by
\begin{eqnarray}
N &=& \sqrt{1-{2M \over r}}~~, ~~~b(r) = {r_0^2+2M(r-r_0) \over r},~~
\omega = {2J \over r^3} = {2aM^2 \over r^3 }~~,~~~K=1. \label{e2}
\end{eqnarray}
Here $a=J/M^2$ is the spin parameter. This wormhole has mass $M$ and a throat at $r_0 \ge 2M \ge 0$.

An ergoregion may be present around the throat of a rotating wormhole. This region can be determine when $g_{tt}=-(N^2-\omega^2 r^2 \sin^2 \theta) \geq 0$ and the ergosurface by $g_{tt}=0$. The ergosurface for the metric \eqref{e1} is given by
\begin{equation}
N^2-\omega^2 r^2 K^2 \sin^2 \theta = 0. \label{e3}
\end{equation}
Since the ergoregion doesn't extent up to the poles $\theta=0$ and $\theta=\pi$, there exist a critical angle $\theta_c$,  where the ergosphere exists in between $\theta_c$ and $\pi - \theta_c$,  for all $0< \theta_c \leq \pi/2$. This critical angle can determined at the throat of the wormhole using eq. \eqref{e3} as
\begin{equation}
\sin \theta_c = \left|{N_0 \over \omega_0 r_0 K_0} \right|
\end{equation}
Further, the ergosphere will only exist when the spin parameter $a$ crosses a critical limit $a_c$,  corresponding to $\sin \theta_c = 1$ or $\omega_c=N_0/r_0 K_0$. For the wormhole metric \eqref{e2} with $M=1$ and $r_0=2.2$, the critical value  {is} $a_c=0.729657$. The ergosphere for these values can be seen in Fig. \ref{f1}.

\section{ Propagation of light in a rotating wormhole spacetime}

When a wormhole is between a  star (bright source of light)  and a viewer, the light comes  to the viewer after being  deviated by the gravitational field of wormhole. However, certain portion of the photons (with small impact parameters) emanated by the star  end  up falling into the wormhole, i.e. not able to  reach the observer, giving as a result,  a dark sector in the sky dubbed as shadow.  The boundary of the shadow apparently is determined by unstable circular photon orbits. In order to find the unstable orbit, one is required  to study the geodesic configuration.

The motion of a photon in the rotating wormhole  spacetime (\ref{e1})   can be described by the Lagrangian,   (${\textbf{\L}} = \frac{1}{2} g_{\mu \nu}  \dot{x}^\mu \dot{x}^\nu$) as:
\begin{equation}
{\textbf{2\L}}= - N^2\dot{t}^2 +  \frac{ \dot{ r}^2}{1-b/r} + r^2K^2[ ~{\dot {\theta}}^2+ \sin^2 \theta (\dot \phi  - \omega \dot t)^2~].
\end{equation}
Here an overdot denotes the  derivative  with respect to the affine parameter $\lambda$. Here  the Lagrangian does not depend on $t$ and $\phi$,  as a result,  we have only two constants of motion,  which are the energy $E$ and the angular momentum $L$  in the direction of the axis of symmetry  of the photon.
\begin{eqnarray}
p_t &=& \frac{\partial{\textbf{\L}}}{\partial \dot{t}}=-N^2\dot{t}-\omega r^2K^2\sin^2\theta(\dot{\phi}-\omega\dot{t})=-E, \nonumber \\
p_\phi &=& \frac{\partial{\textbf{\L}}}{\partial \dot{\phi}}= r^2K^2\sin^2\theta(\dot{\phi}-\omega\dot{t})=L. \nonumber
\end{eqnarray}
These  two equations yield:
\begin{equation}
\dot{t}=\frac{E-\omega L}{N^2},~~~ \dot{\phi}=\frac{L}{r^2K^2\sin^2\theta}+\frac{\omega(E-\omega L)}{N^2}.
\end{equation}
Now, we calculate the $r$  and $\theta$-component of the momentum as,
\begin{equation}
p_r=\frac{\partial\textbf{\L}}{\partial \dot{r}}=\frac{\dot{r}}{1-\frac{b}{r}} ,~~~ p_\theta=\frac{\partial\textbf{\L}}{\partial \dot{\theta}}=r^2K^2\dot{\theta}. \label{e7}
\end{equation}
To achieve a general formula for finding the contour of a shadow we use  Hamilton-Jacobi method  for the null geodesic equations in the general rotating spacetime \eqref{e1}. The Hamilton-Jacobi method  determines the geodesics  via the equation
\begin{equation}
\frac{\partial S}{\partial \lambda} = -\frac{1}{2}g^{\mu\nu}\frac{\partial S}{\partial x^\mu}\frac{\partial S}{\partial x^\nu}, \label{e8}
\end{equation}
where $\lambda$ is the affine parameter along the geodesics, $g_{\mu \nu} $ are the components of the metric tensor, and $S$ is the Jacobi action with the following separable ansatz,
\begin{equation}
S = \frac{1}{2}\mu^2\lambda-Et+L\phi+S_r(r)+S_\theta(\theta), \label{e9}
\end{equation}
where  $S_r$  and $S_\theta $ are the functions of $r$ and $\theta$, respectively and  $\mu$ is the mass of the test particle.  For photons, $\mu =0$. Here the rotating wormhole  metric functions $N,~ b,~ K$ and  $\omega$,  depend only on the radial coordinate, therefore, following \cite{Ned},  the Hamilton-Jacobi equation will be  separable.  Inserting Eq. \eqref{e9} into Eq. \eqref{e8}  one can obtain  the following equations for the functions $S_r(r)$  and $S_\theta(\theta)$ \cite{Ned}:

\begin{eqnarray}
&& \hspace{-0.5cm} \left(\frac{d S_\theta}{d \theta}\right)^2 = Q-\frac{L^2}{\sin^2\theta}, \label{e10}\\
&& \hspace{-0.5cm} \left(1-\frac{b}{r}\right)N^2\left(\frac{d S_r}{d r} \right)^2 = (E-\omega L)^2-\Big(\mu^2N^2
   +Q\frac{N^2}{r^2K^2}\Big), \label{e11}
\end{eqnarray}
where $Q$ is the Carter constant.\\

We know $p_r  =  \frac{\partial S}{\partial r}  = \frac{dS_r}{d r} $  and $  p_\theta  = \frac{\partial S}{\partial  \theta}  = \frac{dS_\theta}{d  \theta}  $, therefore, equations \eqref{e7}, \eqref{e10}  and \eqref{e11} yield \cite{Ned},
\begin{equation}
\frac{N}{\sqrt{1-b/r}}\frac{dr}{d\lambda}=\pm\sqrt{R(r)},~~~  r^2K^2\frac{d\theta}{d\lambda}=\pm\sqrt{T(\theta)}, \label{e12}
\end{equation}
where,
\begin{eqnarray}
T(\theta) &=& Q-\frac{L^2}{\sin^2\theta},\\
R(r) &=& (E-\omega L)^2-\left(\mu^2N^2+Q\frac{N^2}{r^2K^2}\right).
\end{eqnarray}
Hence, the Jacobi action assumes the following form,
\begin{equation}
S = \frac{1}{2}\mu^2\lambda-Et+L\phi+\int\sqrt{\frac{R(r)}{N^2(1-b/r)}}~dr+\int\sqrt{T(\theta)}~d\theta.
\end{equation}

The geodesic  equations  in stationary and axisymmetric spacetimes are parameterized by the constants of motions $E, ~L$ and $Q$. However, the geodesic motion of a photon is described by two independent parameters defined by:
\[   \xi = \frac{L}{E}  , ~~~ \eta = \frac{Q}{E^2}.  \]
Here, $\xi$  and $\eta$ are known as impact parameters  and a new affine parameter
$\widetilde{\lambda} = \lambda E $. After  eliminating  the energy from the geodesic equations,    the path    of  the photon is parameterized only by  $\xi   $ and $\eta$. One can write   the functions $R(r)$ and $ T(\theta)$ in terms of the impact parameters as,
\begin{eqnarray}
R(r) &=& (1-\omega\xi)^2-\eta\frac{N^2}{r^2K^2} ,\\
T(\theta) &=& \eta-\frac{\xi^2}{\sin^2\theta}.
\end{eqnarray}
For a photon $\mu =0$.

\section{Wormhole shadow}
\begin{figure}
\includegraphics[scale=0.31]{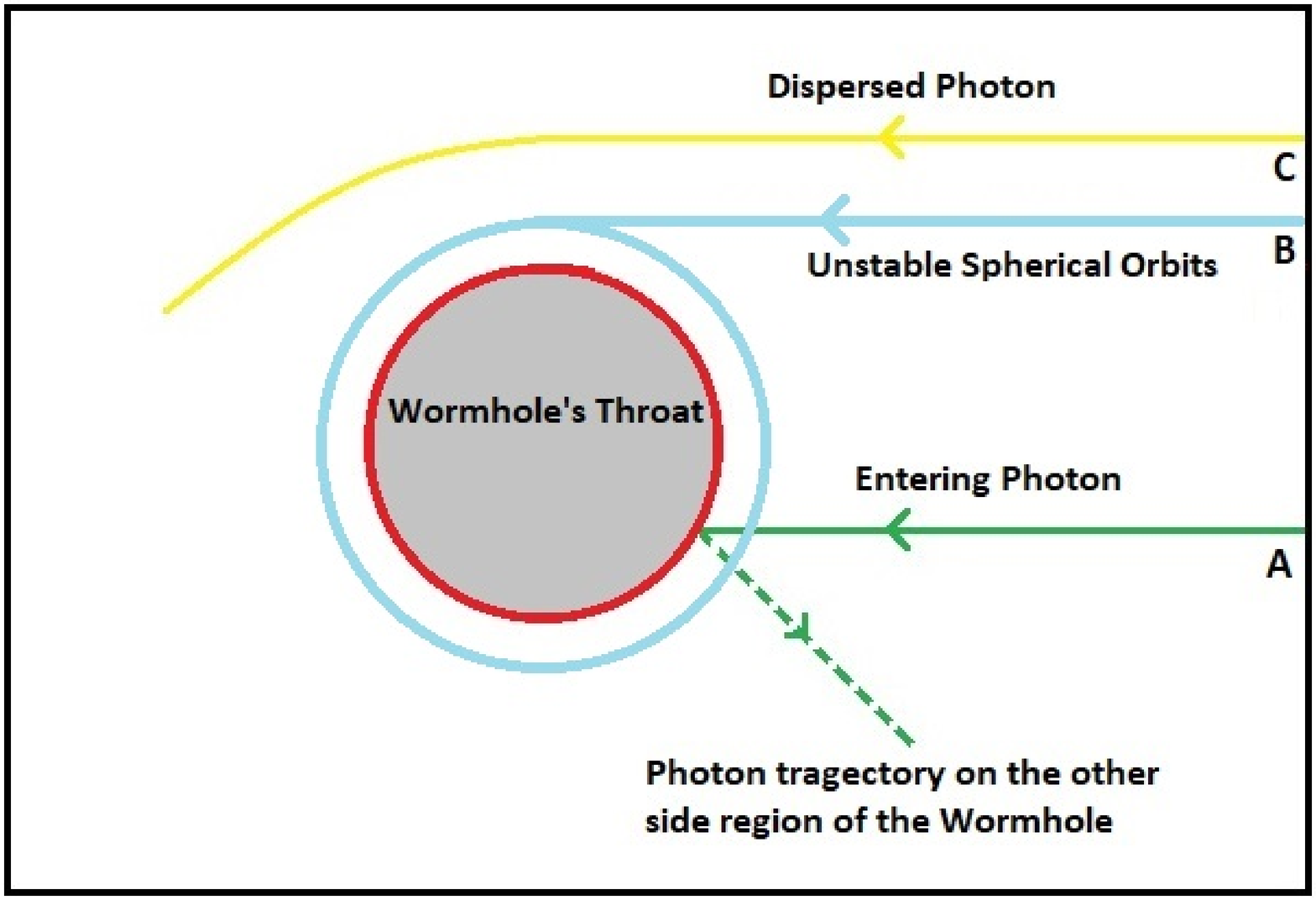}
\caption{Schematic diagram of the photon orbits around the wormholes.}
\label{f2}
\end{figure}

Two regions of spacetime are joined through wormhole (a tunnel-like structure, without any horizon or singularity within it).  Let us  assume that  the   photons  are coming from a star   towards the wormhole,  (placed between the observer and the star)  and illuminate one region,  and no light sources are remaining in the neighborhood of the throat in the other region. In the first region,  the possible orbit  of the photons  around the wormhole are as follows:  (i) orbits entering  into the wormhole and passing through its throat  (ii)    dispersed  from  the wormhole to infinity.  A remote viewer  will be  capable of   observing  only dispersed  photons   from the wormhole (see figure \ref{f2}). However the photon apprehended by the wormhole will form a dark patch. This dark region detected on the bright background is known as  shadow of the wormhole.  To find the wormhole shadow,  i.e. dark region in the  viewer's atmosphere in the presence of a  shiny  background, one would have to  start   finding  out the critical orbits that separate the evading and falling photons. Actually our prime  motive is to estimate these critical geodesics or unstable circular orbits. In order to acquire the boundary of the wormhole shadow,  one needs to study  the radial motion of photons around the wormholes.  The critical orbits (described by certain critical values of the impact parameters  $\xi$ and $\eta$ )   distinguish the dispersed and apprehended photons. A slight disturbance  in these critical values can change  it either to a seepage or to a apprehend orbit.  As a result, the critical impact parameters describe  the boundary of a shadow. Thus  the boundary of the shadow is determined by the unstable circular photon orbits. By studying the radial geodesic equation, one can calculate the critical orbit, which can be expressed  in the form of an energy conservation equation,
\begin{equation}
\left(\frac{dr}{d\bar{\lambda}}\right)^2 +V_{eff}=1 ,~~ V_{eff}=1-\frac{1}{N^2}\left(1-\frac{b}{r}\right)R(r),
\end{equation}
where $V_{eff}$   is the effective potential that describes  the geodesic motion of a photon around the wormhole.  The null geodesics are divided into two classes depending on their impact parameters: (i)  trajectories of photon   passing  through the wormhole (ii)  orbits of the photon   dispersing  to infinity.   The boundary among  the two classes is destined  through a family of unstable spherical orbits  fulfilling  the following conditions,
\begin{equation}
V_{eff}=1,~~~ \frac{V_{eff}}{dr}=0,~~~ \frac{d^2V_{eff}}{dr^2}\leq 0 . \label{e19}
\end{equation}

 \begin{figure}
\includegraphics[scale=0.35]{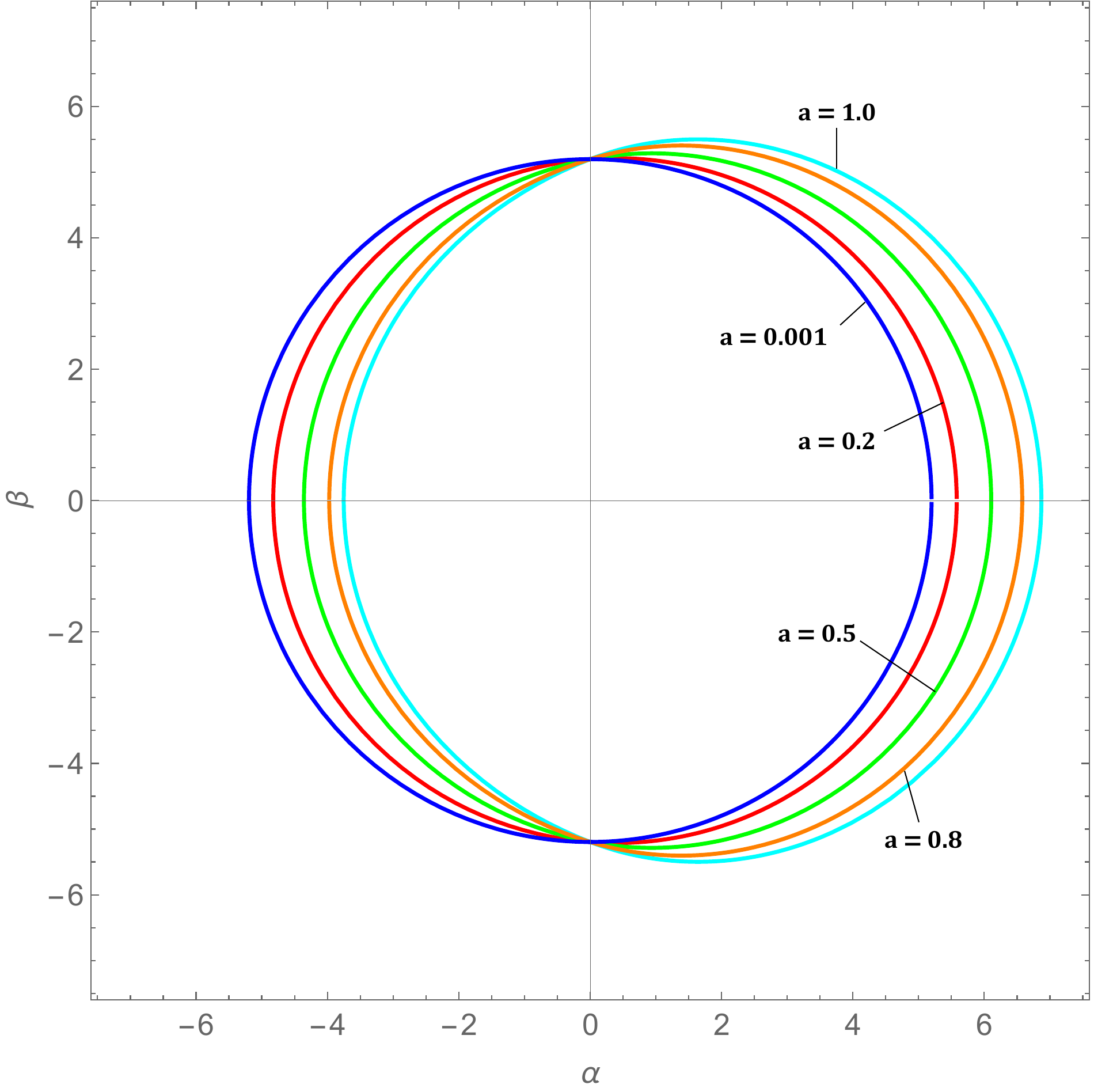}
\caption{Shadow of a wormhole for different $a$ and $r_0=2.5$. The axes are in units of $M$.}
\label{fig3}
\end{figure}

\begin{figure}
\includegraphics[scale=0.4]{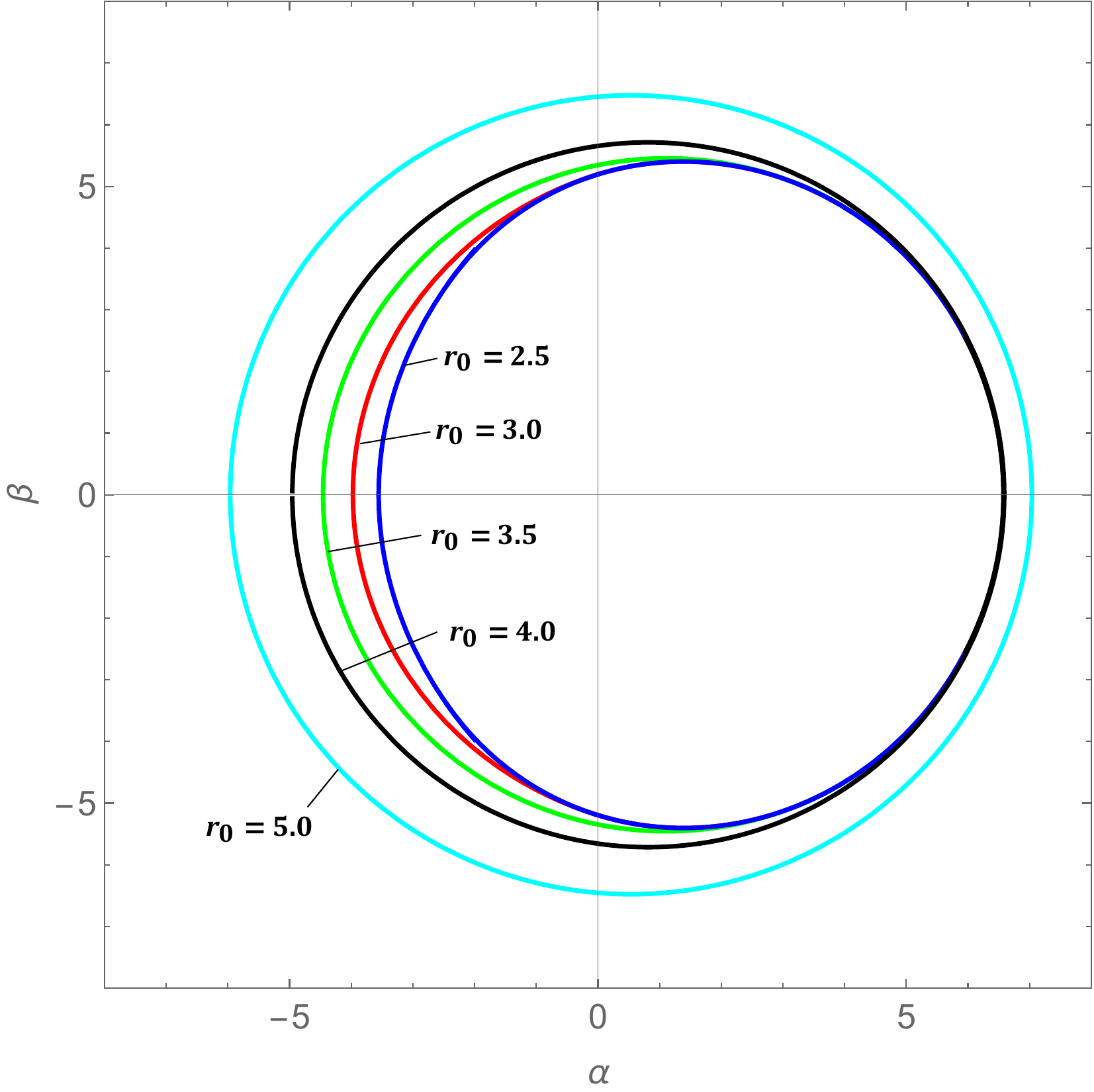}
\caption{Shadow of a wormhole for different $r_0$ and $a=0.8$. The axes are in units of $M$.}
\label{fig4}
\end{figure}

The first condition comes from the fact that the radial motion has a turning point $dr/d\bar{\lambda} = 0  $, when the photon will scatter from  the  wormhole.   The last  conditions  are due to the maximum of the effective potential  for the critical orbit between seepage and plunge motion.

While  writing down the above conditions in terms of $R(r)$, one could assume that the functions $N$ and $ (1-\frac{b}{r})$  are finite and non-zero outside the throat of the wormhole. Therefore, for the unstable circular orbits lying outside the throat, i.e., having radii greater than $r_0$, the above set of conditions can be written in terms of $R(r)$ as,
\begin{equation}
R(r)=0 ,~~~ \frac{dR}{dr}=0,~~~ \frac{d^2R}{dr^2}\geq 0.
\end{equation}
Using the  equations  $ R(r)=0 , ~~ dR/dr=0$, one can easily find two  expressions  for the impact parameters $\xi$ and $\eta$ in terms of the radial coordinate  of the unstable circular orbit as,
\begin{eqnarray}
\eta &=& \frac{r^2K^2}{N^2}(1-\omega\xi)^2 , \label{e21}\\
\xi &=& \frac{\Sigma}{\Sigma\omega-\omega^{'}},~~~ \Sigma=\frac{1}{2}\frac{d}{dr}\ln\left(\frac{N^2}{r^2K^2}\right). \label{e22}
\end{eqnarray}
Here prime denotes the differentiation with respect to radial coordinate $r$. This process of finding the location of the unstable spherical orbits in the impact parameter space as given in equations \eqref{e21} and \eqref{e22} was originally derived in \cite{Ned}. However, it has been shown that a wormhole throat can act as a natural location of unstable circular orbits \citep{RS2019}. For such unstable orbits, $ (1-b/r)$ vanishes and hence, $V_{eff}=1$ is automatically satisfied. Therefore, for unstable circular orbits located at the throat, the second conditions in \eqref{e19} can be reduced to \citep{WS4},
\begin{equation}
(1-\omega_0\xi)^2-\eta\frac{N_0^2}{r_0^2K_0^2}=0,\label{e23}
\end{equation}
where the subscript `0' implies that the functions are evaluated at the throat.

\begin{figure*}
\includegraphics[scale=0.5]{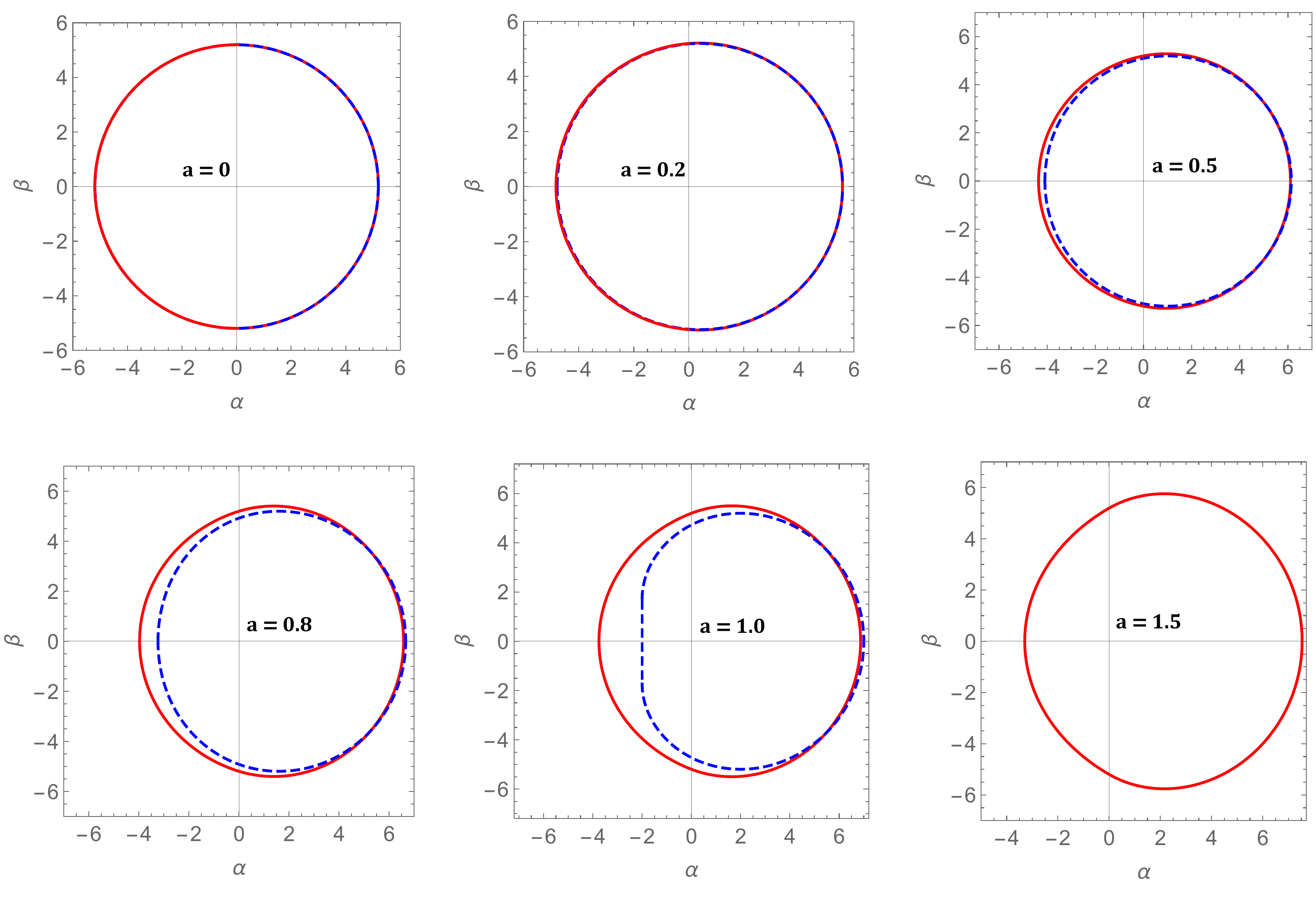}
\caption{Shadow of a wormhole (red solid curve) and a Kerr black hole (blue dashed curve) for different $a$. Here $r_0=3.0$. The axes are in units of $M$.}
\label{fig5}
\end{figure*}

For the set of unstable photon orbits, equations \eqref{e21}, \eqref{e22} and \eqref{e23} express the critical locus of the impact parameters. In other words, these equations  express  the boundary of the shadow in the impact parameter space. However, for true observation, the remote viewer will see  the apparent shape of a shadow in his sky (a plane passing through the center of the wormhole and normal to the line linking it with the viewer). The celestial coordinates connected to the actual astronomical measurements that span a two-dimensional plane are defined as \cite{vaz},
\begin{eqnarray}
\alpha &=& \lim_{r_o\rightarrow\infty}\left(-r_o^2\sin\theta_o \left[\frac{d\phi}{dr}\right]_{(r_o,\theta_o)}\right),\\
\beta &=& \lim_{r_o\rightarrow\infty}\left(r_o^2  \left[\frac{d\theta}{dr}\right]_{(r_o,\theta_o)}\right),
\end{eqnarray}
where $r_o$ is  the position coordinate  of the remote viewer taken very large (far away from the wormhole) and $\theta_o$ is  the angular coordinate. It is actually the angle of inclination between the axis of symmetry of the wormhole and the direction to the viewer.

\begin{figure*}
\includegraphics[scale=0.5]{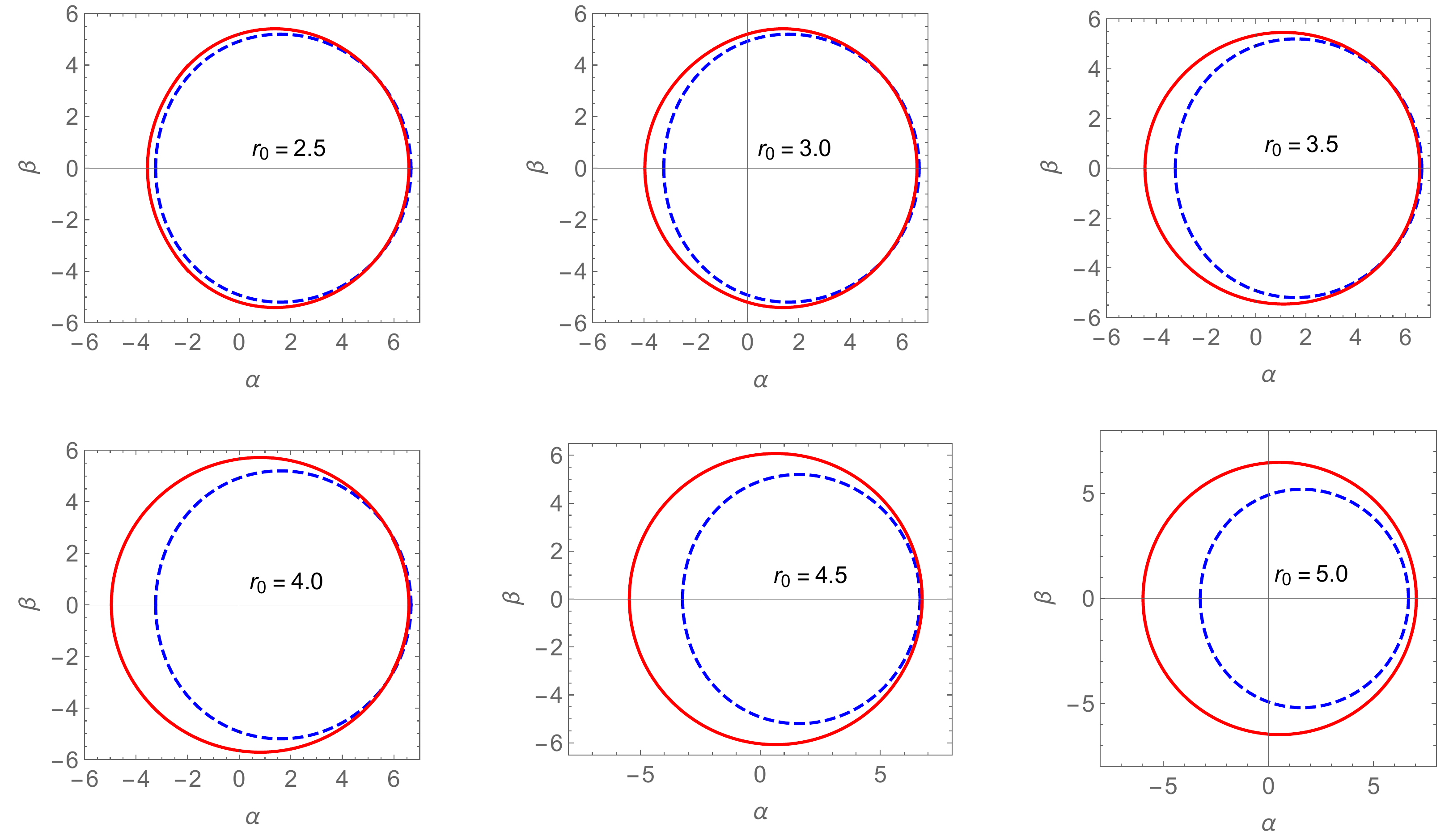}
\caption{Shadow of a wormhole (red solid curve) and a Kerr black hole (blue dashed curve) for different $r_0$. Here $a=0.8$. The axes are in units of $M$.}
\label{fig6}
\end{figure*}

The celestial coordinates  $\alpha$ and $\beta$  are  the  discernible perpendicular distances of the shadow  image, as observed from the rotation axis and    from its projection on the equatorial plane respectively. Using the geodesic equations  and values of the four-velocity components and after some simple algebraic manipulation, the  celestial coordinates assume the following forms:
\begin{eqnarray}
\alpha &=& -\frac{\xi}{\sin\theta_0} , \label{e26}\\
\beta &=& \left(\eta-\frac{\xi^2}{\sin^2\theta_0}\right)^{1/2}. \label{e27}
\end{eqnarray}
After knowing  the expressions of celestial coordinates and impact parameters, one can construct the shadow of wormholes. In the $(\alpha,\beta)$-plane, Eqs. \eqref{e21}, \eqref{e22}, \eqref{e26} and \eqref{e27} define the part of the shadow boundary formed by the unstable circular orbits which lie outside the throat. However, the part of the shadow boundary formed due to the unstable circular orbits which are located at the throats, we obtain, from Eqs. \eqref{e23}, \eqref{e26} and \eqref{e27},
\begin{eqnarray}
(N_0^2-\omega_0^2 r_0^2K_0^2\sin^2\theta_{obs})\alpha^2-2\omega_0 r_0^2K_0^2\sin\theta_{obs} \,\alpha -r_0^2K_0^2+N_0^2\beta^2=0.\label{e28}
\end{eqnarray}
Therefore, the complete contour of the shadow is given by the combination of Eq. \eqref{e26}-\eqref{e27} [with $\xi$ and $\eta$ given by Eq. \eqref{e21} and \eqref{e22}] and Eq. \eqref{e28} \citep{WS4}.

Figures \ref{fig3} and \ref{fig4} show the shadow the wormhole given above for different values of the spin and throat size. Note that, for a given wormhole size $r_0$, the shadow shape deviates from circularity as we increase the spin $a$ (see fig. \ref{fig3}). On the other hand, for a given spin $a$, the shadow becomes more and more circular as we increase the wormhole size $r_0$ (see fig. \ref{fig4}). We also have compared our results with those of a Kerr black hole in figs. \ref{fig5} and \ref{fig6}. Note that, for small spin and smaller wormhole size, its shadow mimics those of the black hole. However, with increasing either the spin or the throat size, the shadow of a wormhole start deviating from that of a Kerr black hole. Detection of such deviation may possibly indicate the presence of a wormhole.

\section{Constraining the wormhole parameters using the M87$^*$ results}

We now constrain the size and the spin of the wormhole using the results from M87$^*$ observation \cite{telescope}. For this purpose, we use the average angular size of the shadow and its deformation from circularity. Since the shadow has reflection symmetry with respect to the $\alpha$-axis, its geometric center $(\alpha_c,\beta_c)$ is given by $\alpha_c=1/A\int \alpha dA$ and $\beta_c=0$, $dA$ being an area element. We first define an angle $\phi$ between the $\alpha$-axis and the vector connecting the geometric centre $(\alpha_c,\beta_c)$ with a point $(\alpha,\beta)$ on the boundary of a shadow. Therefore, the average radius $R_{av}$ of the shadow is given by. \cite{Bambi}
\begin{equation}
R_{av}^2=\frac{1}{2\pi}\int_{0}^{2\pi}l^2(\phi)\; d\phi,
\end{equation}
where $l(\phi)=\sqrt{[\alpha(\phi)-\alpha_c]^2+\beta(\phi)^2}$ and $\phi=\tan^{-1}[\beta(\phi)/(\alpha(\phi)-\alpha_c)]$. Following \cite{telescope}, we define the deviation $\Delta C$ from circularity as,
\begin{equation}
\Delta C=\frac{1}{R_{av}}\sqrt{\frac{1}{2\pi}\int_{0}^{2\pi}(l(\phi)-R_{av})^2\; d\phi}.
\end{equation}
Note that $\Delta C$ is the fractional RMS distance from the average radius of the shadow.

\begin{figure}[h]
\centering
\includegraphics[scale=0.85]{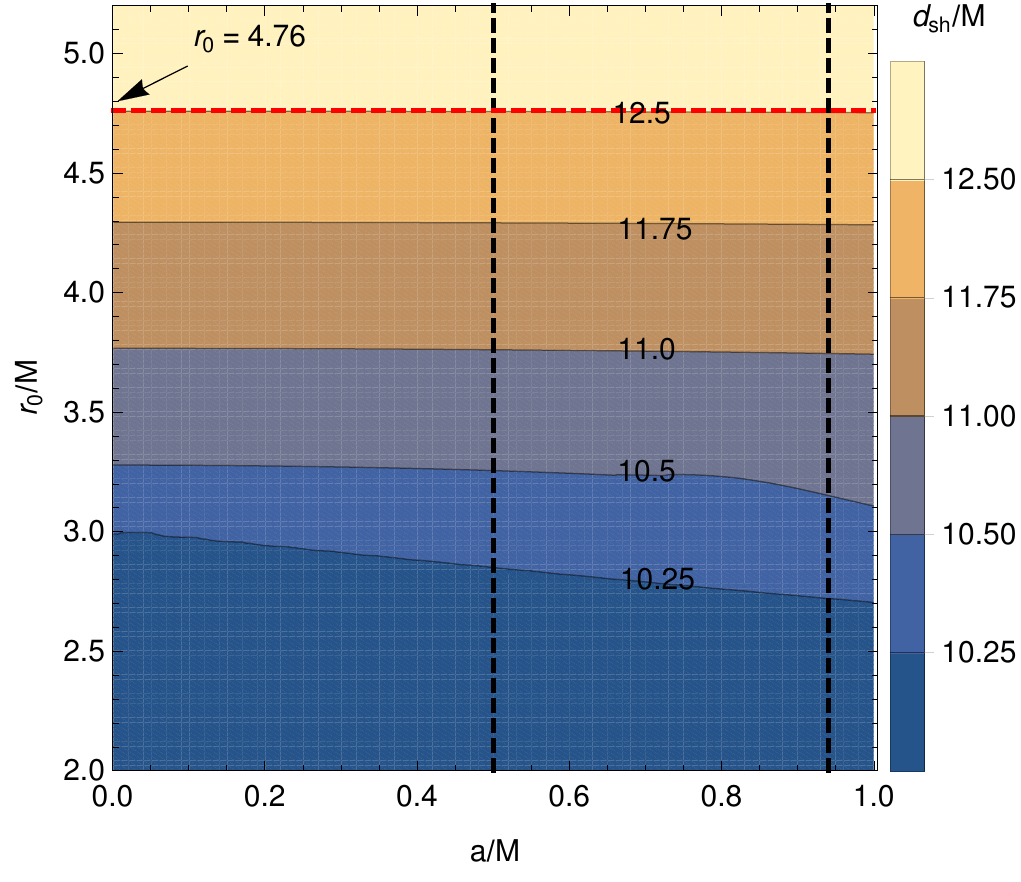}
\includegraphics[scale=0.85]{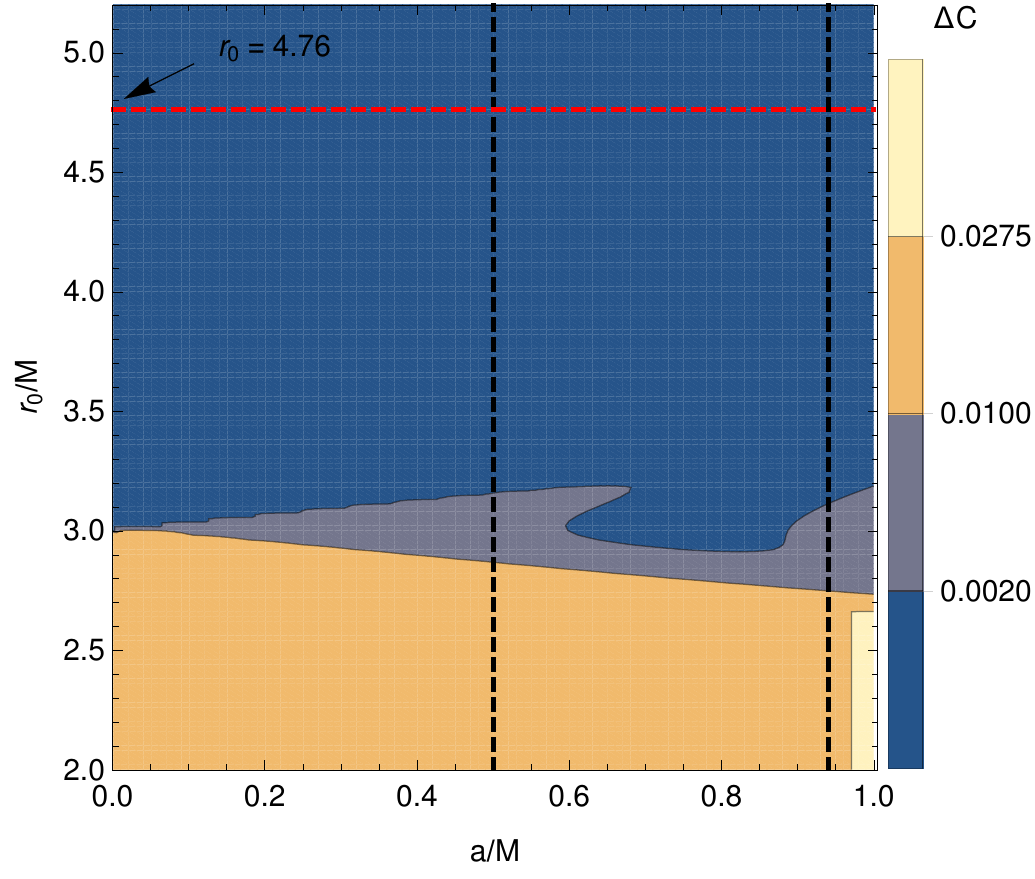}
\caption{Dependence of the angular size and the deviation of the shadow on $r_0$ and spin. The region between the two black dashed vertical lines indicate the spin range $0.5\leq a_{*}\leq 0.94$. The horizontal red dashed line indicates the critical throat size $r_{0c}=4.76M$, below which both the size and deviation from circularity of the shadow is consistent with the M87$^*$ results.}
\label{fig:constraint}
\end{figure}

According to EHT collaboration \cite{telescope}, the angular size of the observed shadow is $\Delta\theta_{sh}=42\pm 3$ $\mu$as, and the deviation $\Delta C$ is less than $10\%$. Also, following \cite{telescope}, we take the distance to M87$^*$ to be $D=(16.8\pm 0.8)$ Mpc and the mass of the object to be $M=(6.5\pm 0.7)\times 10^9 M_\odot$. These numbers imply that the average diameter of the shadow should be,
\begin{equation}
\frac{d_{sh}}{M}=\frac{D\Delta\theta_{sh}}{GM}=11.0\pm 1.5,
\end{equation}
where the errors have been added in quadrature. The above quantity must be equal to $\frac{2R_{av}}{M}$. In Fig. \ref{fig:constraint}, we have shown the average diameter and the deviation from circularity of the shadow for different values of the spin and the wormhole throat size. Here, we have taken the inclination angle to be $\theta_o=17^\circ$, which the jet axis makes to the line of sight \citep{telescope}. Also, according to EHT collaboration, the spin lies within the range $0.5\leq a_{*}\leq 0.94$, where $a_{*}=a/M$. Note that, for this spin range, the diameter of the shadow is always greater than the minimum allowed value of $9.5$. However, the diameter will be within the maximum allowed value $12.5$ if the wormhole throat size is restricted below some critical value $r_0\leq r_{0c}=4.76 M$. Note that the deviation from circularity is always less than $2.75\%$, i.e., $\Delta C\leq 0.0275$ for the allowed throat size $r_0<r_{0c}=4.76M$ and spin range $0.5\leq a_{*}\leq 0.94$.

Here we would like to mention that, using integrable geodesic equations, shadows of rotating traversable wormhole geometries which belong to the class same as ours but with different forms of the metric functions were successfully studied in many prior works like references \cite{Ned}, \cite{WS3}, \cite{WS4}, and similar conclusions about the dependency of the shadow boundary on the wormhole spin were obtained. However, these works did not considered constraining the wormhole parameters using observational data. In our work here, along with studying the dependence of the shadow on the wormhole spin as well as on the wormhole throat size, we have constrained the parameters using the M87$^*$ results.

\section{Concluding Remarks}
In this paper, we have investigated the shadow of a rotating traversable wormhole and compared our results with those of Kerr black hole of same mass and spin. The results obtained here will have several implications in the study of shadow. The throat of a wormhole plays very crucial role in shadow formation. We have found that, for small spin and smaller wormhole throat size, the shadow of a wormhole mimics those of the black hole. However, with increasing either the spin or the throat size, the shadow of a wormhole start deviating from that of a black hole. Detection of such deviation may possibly indicate the presence of a wormhole. { In section V, we have constrained the size and the spin of the wormhole, using the results from M87$^*$ observation \cite{telescope}. In Fig. \ref{fig:constraint}, we have shown the average diameter and the deviation from circularity of the shadow for different values of the spin and the wormhole throat size, taking inclination angle to be $\theta_o=17^\circ$, which the jet axis makes with the line of sight. Note  that, the spin lies within the range $0.5\leq a_{*}\leq 0.94$, where $a_{*}=a/M$, and the diameter of the shadow is always greater than the minimum allowed value of $9.5$. However, the diameter will be within the maximum allowed value $12.5$, if the wormhole throat size is restricted below some critical value $r_0\leq r_{0c}=4.76 M$. Also, the deviation from circularity is always less than $2.75\%$, i.e., $\Delta C\leq 0.0275$ for the allowed throat size $r_0<r_{0c}=4.76M$ and spin range $0.5\leq a_{*}\leq 0.94$.}  In other words, the results obtained here indicate that a wormhole having reasonable spin or throat size, can be distinguished from a black hole through observations of their shadow. Also worth noting a recent paper by Gralla et al. \cite{Gralla} in which the emission profile originating near a black hole has been discussed and the "photon rings" surrounding the dark black hole shadow has been studied, including the implications of the recent M87* Event Horizon Telescope observations and mass measurement on it. Previous studies claimed that the observed emission will peak near the photon ring, but this paper suggested otherwise. Distinguishing between a "photon ring" (light rays that complete at least $n=5/4$ orbits) and a "lensing ring" (light rays that complete between $3/4$ and $5/4$ orbits); it showed that, for optically thin emission, "photon ring" produced a sharp feature near the critical impact parameter but the peak is so narrow, and the brightness being logarithmic, it never makes a significant contribution to the observed flux. This interesting study may motivate future studies on the emission spectra of a accretion disk surrounding the wormhole described in this paper. \\


\section*{Acknowledgments}

FR would like to thank the authorities of the Inter-University Centre for Astronomy and Astrophysics, Pune, India for providing research facilities.
This work is a part of the project submitted in DST-SERB, Govt. of India.


\end{document}